\documentclass[sigconf,nonacm]{acmart}

\renewcommand\footnotetextcopyrightpermission[1]{} % removes footnote with conference information in first column
\usepackage{graphicx}
\usepackage[
font=Large
]{subfig}
\usepackage{tikz}
\usetikzlibrary{shapes, arrows,petri,topaths,backgrounds,patterns,positioning}
\usepackage{pgfplots}
\usepgfplotslibrary{statistics}
\usepackage{units}
\usepackage{booktabs}

\definecolor{amber}{rgb}{1.0, 0.75, 0.0}
\definecolor{awesome}{rgb}{1.0, 0.13, 0.32}
\definecolor{ao(english)}{rgb}{0.0, 0.5, 0.0}

\pgfplotsset{
	colormap={blackwhite}{gray(0cm)=(1.0); gray(1cm)=(0.0001)},
}

\pgfplotsset{
	cycle list/.define={my marks}{
		every mark/.append style={solid,fill=\pgfkeysvalueof{/pgfplots/mark list fill}},very thick\\
		every mark/.append style={solid,fill=\pgfkeysvalueof{/pgfplots/mark list fill}},very thick\\
		every mark/.append style={solid,fill=\pgfkeysvalueof{/pgfplots/mark list fill}},very thick\\
		every mark/.append style={solid,fill=\pgfkeysvalueof{/pgfplots/mark list fill}},very thick\\
		every mark/.append style={solid,fill=\pgfkeysvalueof{/pgfplots/mark list fill}},very thick\\
		every mark/.append style={solid,fill=\pgfkeysvalueof{/pgfplots/mark list fill}},very thick\\
		every mark/.append style={solid,fill=\pgfkeysvalueof{/pgfplots/mark list fill}},very thick\\
		every mark/.append style={solid,fill=\pgfkeysvalueof{/pgfplots/mark list fill}},very thick\\
		every mark/.append style={solid,fill=\pgfkeysvalueof{/pgfplots/mark list fill}},very thick\\
		every mark/.append style={solid,fill=\pgfkeysvalueof{/pgfplots/mark list fill}},very thick\\
		every mark/.append style={solid,fill=\pgfkeysvalueof{/pgfplots/mark list fill}},very thick\\
		every mark/.append style={solid,fill=\pgfkeysvalueof{/pgfplots/mark list fill}},very thick\\
		every mark/.append style={solid,fill=\pgfkeysvalueof{/pgfplots/mark list fill}},very thick\\
		every mark/.append style={solid,fill=\pgfkeysvalueof{/pgfplots/mark list fill}},very thick\\
		every mark/.append style={solid,fill=\pgfkeysvalueof{/pgfplots/mark list fill}},very thick\\
		every mark/.append style={solid,fill=\pgfkeysvalueof{/pgfplots/mark list fill}},very thick\\
	},
}

\AtBeginDocument{%
	\providecommand\BibTeX{{%
			\normalfont B\kern-0.5em{\scshape i\kern-0.25em b}\kern-0.8em\TeX}}}

\usepgfplotslibrary{groupplots}
\usetikzlibrary{pgfplots.groupplots}
\usetikzlibrary{pgfplots.colorbrewer}

\begin{document}
\title{Twitch Gamers: a Dataset for Evaluating Proximity Preserving and Structural Role-based Node Embeddings}

\author{Benedek Rozemberczki}
\affiliation{%
  \institution{The University of Edinburgh}
  \city{Edinburgh}
  \country{United Kingdom}}
\email{benedek.rozemberczki@ed.ac.uk}

\author{Rik Sarkar}
\affiliation{%
  \institution{The University of Edinburgh}
  \city{Edinburgh}
  \country{United Kingdom}}
\email{rsarkar@inf.ed.ac.uk}

\begin{abstract}
Proximity preserving and structural role-based node embeddings have become a prime workhorse of applied graph mining. Novel node embedding techniques are often tested on a restricted set of benchmark datasets. In this paper, we propose a new diverse social network dataset called \textit{Twitch Gamers} with multiple potential target attributes. Our analysis of the social network and node classification experiments illustrate that \textit{Twitch Gamers} is suitable for assessing the predictive performance of novel proximity preserving and structural role-based node embedding algorithms.  
\end{abstract}

\maketitle
\thispagestyle{empty}
\section{Introduction}\label{sec:twitch_introduction}
The prediction of unknown node attributes using vertex features is a central problem in both theoretical and applied graph mining research. One way to create high quality node features is to embed the vertices in an Euclidean space. Node embedding algorithms are frequently used as an upstream unsupervised feature extraction method to distill useful features for downstream supervised models. Their success is mainly due to  the favorable algorithmic qualities they have such as runtime and memory efficiency. In addition to efficiency, the extracted node representations are known to be robust to hyperparameter changes \cite{deepwalk,perozzidontwalk,rozemberczki2019gemsec} and the learned features are reusable when new downstream machine learning tasks come up \cite{rolx,rolebased}. Node embedding techniques are typically evaluated on a limited number of public benchmark datasets \cite{deepwalk,perozzidontwalk, rolx,henderson2011s,tang2015line,netmf}, which are not compatible with newly proposed attribute based algorithms~\cite{musae,rozemberczki2019gemsec,feather}. This highlights the need for new benchmark datasets which are rich in attributes.

\textbf{Present work.} In order to foster node embedding research we publicly release \textit{Twitch Gamers}: a medium sized undirected social network of online streamers with multiple interesting vertex attributes. Using \textit{Twitch Gamers},  the predictive performance of a node embedding algorithm can be tested on multiple new challenging node classification and vertex level regression problems. Potential machine learning tasks include the identification of dead accounts,  selection of users that stream explicit content and broadcaster language prediction. Our work creates opportunity for the assessment of numerous existing node representation learning techniques and newly developed vertex embedding procedures.

\textbf{Main contributions.} The most important contributions of our paper can be summarized as follows:

\begin{enumerate}
    \item We release \textit{Twitch Gamers}: a new social network dataset which we specifically collected for benchmarking the vertex classification performance of proximity preserving and structural role-based node embedding techniques.
    \item We carry out a descriptive analysis of the social network and underlying generic vertex features and argue that it is suitable for testing novel node embedding methods.
    \item We evaluate the performance of standard node embedding algorithms under various train/test split regimes.
\end{enumerate}

The rest of our work has the following structure. We overview the related work about node embedding procedures in Section \ref{sec:twitch_related_work}. We discuss in Section \ref{sec:dataset} the data collection and the dataset itself. We perform descriptive analysis of the social network and the generic vertex attributes in Section \ref{sec:twitch_analysis}. In Section \ref{sec:twitch_experiments} we showcase the predictive performance of various well known node embedding techniques on the \textit{Twitch Gamers} dataset. The paper concludes with Section \ref{sec:twitch_conclusions} where we discuss potential future work.
\section{Related Work}\label{sec:twitch_related_work}
Given a graph $G=(V,E)$ node embedding techniques learn a function $f: V\to \mathbb{R}^d$ which maps the nodes $ v\in V$ into a $d$ dimensional Euclidean space. When generic vertex features are not available for node classification, proximity preserving and structural role-based node embedding techniques are suitable for distilling high quality reusable feature sets \cite{rolebased}. We will utilize linear runtime node embedding algorithms to showcase that \textit{Twitch Games} is suitable for multi-aspect testing of feature extraction.

Proximity preserving node embedding algorithms \cite{rozemberczki2019gemsec,deepwalk,node2vec,hope,perozzidontwalk} learn this embedding by preserving a certain notion of proximity in the embedding space such as pairwise truncated random walk transition probabilities.  This way nodes that are close to each other in the graph are also close in the embedding space. Structural role-based node embedding techniques on the other hand preserve structural similarity in the embedding space. Nodes which have similar structural properties such as centrality and transitivity are close to each other in the embedding space \cite{rolx, henderson2011s,graphwave}. 
\section{The Twitch Gamers dataset}\label{sec:dataset}
Twitch is a streaming service where users can broadcast live streams of playing computer games. As users can follow each other there is an underlying social network which can be accessed through the public API. In 2018 April we crawled the largest connected component of this social network with snowball sampling  starting from the user called \textit{Lowko}. The released \textit{Twitch Gamers} dataset is a clean subset of the original social network. We filtered out nodes and edges based on the following principled steps:

\begin{enumerate}
    \item \textbf{No missing attributes.} We only kept nodes that have all of the vertex attributes present.
    \item \textbf{Mutual relationships.} We discarded relationships which are asymmetric and only included mutual edges in the released dataset.
    \item \textbf{Member of the largest component.} We only considered nodes which are part of the largest connected component.
\end{enumerate}
The result of this three step data cleaning process is an undirected, single component social network with approximately 168 thousand nodes and 6.79 million edges. Vertices in this restricted subsample do not have any missing node attributes. We summarized the name, meaning, and type of available generic node attributes in Table \ref{tab:twitch_descriptives}. 

\begin{table}[h!]
{\centering\footnotesize
\begin{tabular}{ccc}
\hline
\textbf{Name} & \textbf{Meaning} & \textbf{Type} \\
\hline
  Identifier   &  Numeric vertex identifier.       &  Index    \\
  Dead Account   &   Inactive user account.      &   Categorical   \\
 Broadcaster Language   &   Languages used for broadcasting.      &    Categorical  \\
  Affiliate Status   &  Affiliate status of the user.       & Categorical    \\
 Explicit Content    &  Explicit content on the channel.       & Categorical\\
 Creation Date   &   Joining date of the user.      & Date \\
 Last Update    &  Last stream of the user.       & Date\\
 View Count  &    Number of views on the channel.     & Count\\
 Account Lifetime &   Days between first and last stream.      & Count\\\hline   
\end{tabular}}
\vspace{2mm}
\caption{The name, meaning and type of vertex attributes in the \textit{Twitch Gamers} dataset.}\label{tab:twitch_descriptives}
\end{table}

Categorical attributes such as \textit{Dead Account, Affiliate Status}, and \textit{Explicit Content} can be used as targets for binary classification, while \textit{Broadcaster Language} can be used for multi-class node classification with more than 20 categories. The vertex attributes \textit{View Count} and \textit{Account Lifetime} can serves as target for count data regression problems at the node level. Various other supervised and unsupervised machine learning tasks can be performed on the dataset such as link prediction and community detection with ground truth labels. The \textit{Twitch Gamers} dataset is publicly available at \url{https://github.com/benedekrozemberczki/datasets}.

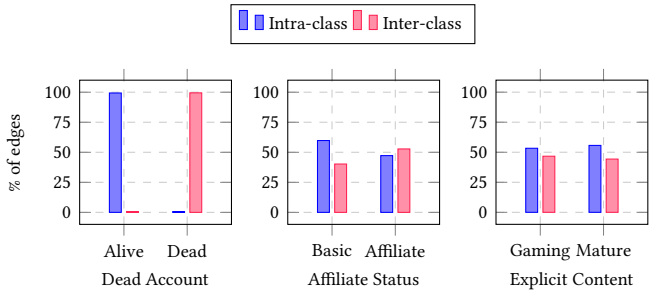
\begin{figure}[h!]
\begin{tikzpicture}
	\tikzset{font={\fontsize{7pt}{8}\selectfont}}
\begin{groupplot}[
	grid=major,
	grid style={dashed, gray!40},
group style={
                      group name=myplot,
                      group size= 3 by 1, horizontal sep=0.75cm,vertical sep=1.5cm},height=3.5cm,width=3.6cm, title style={at={(0.5,1.0)},anchor=south},
                      ylabel style={at={(0.3,0.5)},anchor=south},]
\nextgroupplot[
	legend style={at={(1.80,1.25)},
	legend columns=2,
	anchor=south},
    ybar,
    enlarge x limits=0.70,
    enlarge y limits=0.1,
    ymin=0,
	ylabel=\% of edges,
    ytick={0,25,50,75,100},
    xtick=data,
    ymax=100,
    symbolic x coords={Alive,Dead},
	xlabel=Dead Account,
	bar width=4.5pt,
    ]

\addplot[blue,fill=blue!50] coordinates {
(Alive,99.337)
(Dead,0.548)
};
\addplot[awesome,fill=awesome!50] coordinates {(Alive,0.663)(Dead,99.452)};

\legend{Intra-class, Inter-class}
\nextgroupplot[
	legend style={at={(0.50,1.25)},
	legend columns=2,
	anchor=south},
    ybar,
    enlarge x limits=0.70,
    enlarge y limits=0.1,
    ymin=0,
    ytick={0,25,50,75,100},
	grid=major,
	grid style={dashed, gray!40},
    xtick=data,
    ymax=100,
    symbolic x coords={Basic,Affiliate},
	ylabel=\empty,
	xlabel=Affiliate Status,
	bar width=4.5pt,
    ]

\addplot[blue,fill=blue!50] coordinates {
(Basic,59.781)
(Affiliate,47.223)

};
\addplot[awesome,fill=awesome!50] coordinates {
(Basic,40.219)
(Affiliate,52.777)
};

\nextgroupplot[
	legend style={at={(0.50,1.25)},
	legend columns=2,
	anchor=south},
    ybar,
    enlarge x limits=0.70,
    enlarge y limits=0.1,
    ymin=0,
    ytick={0,25,50,75,100},
    xtick=data,
    ymax=100,
    symbolic x coords={Gaming,Mature},
	ylabel=\empty,
	xlabel=Explicit Content,
	bar width=4.5pt,
    ]

\addplot[blue,fill=blue!50] coordinates {
(Gaming,53.307)
(Mature,55.685)

};
\addplot[awesome,fill=awesome!50] coordinates {
(Gaming,46.693)
(Mature,44.315)
};
\end{groupplot}
\end{tikzpicture}
\caption{The percentage of intra and inter-class edges conditional on the \textit{Dead Account, Affiliate Status}, and \textit{Explicit Content} attributes.}\label{fig:twitch_intra_others}
\end{figure}
\section{Descriptive Analysis}\label{sec:twitch_analysis}
Our descriptive analysis of \textit{Twitch Gamers} focuses on the interaction of graph topology and attributes. Specifically, we investigate which potential target attributes can be predicted well with neighbourhood-preserving and structural role-based techniques.

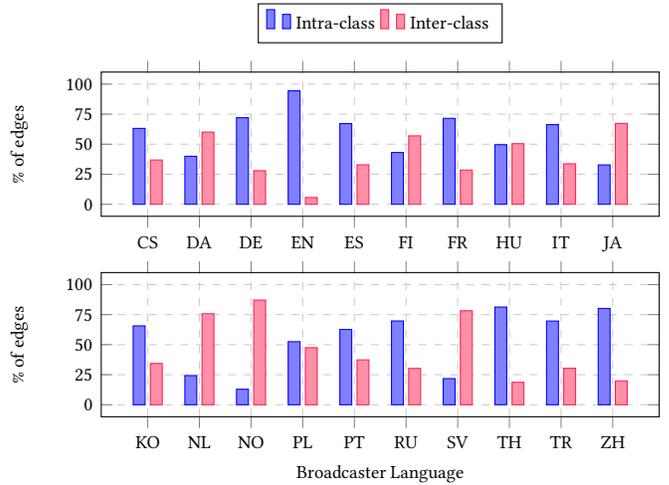
\begin{figure}[h!]
\begin{tikzpicture}
	\tikzset{font={\fontsize{7pt}{8}\selectfont}}
\begin{groupplot}[
	grid=major,
	grid style={dashed, gray!40},
group style={
                      group name=myplot,
                      group size= 1 by 2, horizontal sep=1.25cm,vertical sep=0.75cm},height=3.5cm,width=9.0cm, title style={at={(0.5,1.0)},anchor=south},
                      ylabel style={at={(0.05,0.5)},anchor=south},]
\nextgroupplot[
	legend style={at={(0.50,1.2)},
	legend columns=2,
	anchor=south},
    ybar,
    enlargelimits=0.10,
    ymin=0,
    ytick={0,25,50,75,100},
    xtick=data,
    ymax=100,
    symbolic x coords={CS,DA,DE,EN,ES,FI,FR,HU,IT,JA},
	ylabel=\% of edges,
%	xlabel=Broadcaster language,
	bar width=4.5pt,
    ]

\addplot[blue,fill=blue!50] coordinates {
(CS,63.149)
(DA,39.95)
(DE,72.054)
(EN,94.405)
(ES,67.155)
(FI,43.08)
(FR,71.521)
(HU,49.581)
(IT,66.281)
(JA,32.753)
};
\addplot[awesome,fill=awesome!50] coordinates {(CS,36.851)(DA,60.050)(DE,27.946)(EN,5.595)(ES,32.845)(FI,56.920)(FR,28.479)(HU,50.419)(IT,33.719)(JA,67.247)};
\legend{Intra-class, Inter-class}
\nextgroupplot[
	legend style={at={(0.50,1.25)},
	legend columns=2,
	anchor=south},
    ybar,
    enlargelimits=0.10,
    ymin=0,
    ytick={0,25,50,75,100},
    xtick=data,
    ymax=100,
    symbolic x coords={KO,NL,NO,PL,PT,RU,SV,TH,TR,ZH},
	ylabel=\% of edges,
	xlabel=Broadcaster Language,
	bar width=4.5pt,
    ]

\addplot[blue,fill=blue!50] coordinates {
(KO,65.629)
(NL,24.264)
(NO,12.887)
(PL,52.489)
(PT,62.644)
(RU,69.724)
(SV,21.765)
(TH,81.307)
(TR,69.649)
(ZH,80.168)

};
\addplot[awesome,fill=awesome!50] coordinates {
(KO,34.371)
(NL,75.736)
(NO,87.113)
(PL,47.511)
(PT,37.356)
(RU,30.276)
(SV,78.235)
(TH,18.693)
(TR,30.351)
(ZH,19.832)
};
\end{groupplot}
\end{tikzpicture}

\caption{The percentage of intra and inter-class edges conditional on the \textit{Broadcaster Language} attribute.}\label{fig:twitch_intra_language}
\end{figure}
We plotted the ratio of inter and intra-class edges conditional on the categorical attributes on Figures \ref{fig:twitch_intra_language} and \ref{fig:twitch_intra_others}. These results show that users who broadcast in more commonly spoken language (English, German, French) are more likely to have connections with users who broadcast in the same language. This postulates that proximity preserving node embedding techniques will extract expressive features that can predict \textit{Broadcaster Language} precisely. We also see that Twitch users who churned from the platform are well embedded in the social network and do not form communities. When it comes to the \textit{Affiliate Status} and \textit{Explicit Content} attributes we cannot highlight particular insights about the related linking behaviour of vertices.
\begin{figure}[h!]
	\centering
	\begin{tikzpicture}[scale=0.95,transform shape]
	\tikzset{font={\fontsize{7pt}{8}\selectfont}}
\begin{groupplot}[
	grid=major,
	grid style={dashed, gray!40},
group style={group name=myplot,
                      group size= 2 by 3, horizontal sep=1.25cm,vertical sep=1.15cm},height=2.5cm,width=4.75cm, title style={at={(0.5,0.8)},anchor=south},
                     xlabel style={at={(0.5,-0.25)},anchor=south},
                      ylabel style={at={(0.2,0.5)},anchor=south},]
\nextgroupplot[
    ytick={1,2},
    ylabel=\empty,
    title=\textbf{Dead Account},
    xlabel=$\log$ Degree Centrality,  yticklabels={Alive,Dead},
    ymin=0.5,
    ymax=2.5,
    xmin=-0.5,
    xmax=8.5,
    xtick={0,2,4,6,8}
 ]

    %DONE
    \addplot+[thick,solid,
    black!80,
    fill=cyan!90,mark={diamond*},
    boxplot prepared={
        box extend=0.3, 
lower whisker=0.0,
lower quartile=2.565,
median=3.497,
upper quartile=4.344,
upper whisker=10.471
    },
    ] coordinates {};
    %DONE
    \addplot+[thick,solid,
    black!80,
    fill=cyan!90,mark={diamond*},
    boxplot prepared={
        box extend=0.3, 
lower whisker=0.0,
lower quartile=1.099,
median=2.079,
upper quartile=2.996,
upper whisker=6.182
    },
    ] coordinates {};

\nextgroupplot[
    ytick={1,2},
    ylabel=\empty,
    title=\textbf{Dead Account},
    xlabel=Clustering Coefficient,
    yticklabels={Alive,Dead},
    ymin=0.5,
    ymax=2.5,
    xmin=-0.05,
    xmax=1.05,
    xtick={0,0.25,0.5,0.75,1.0},
    xticklabels={0,0.25,0.5,0.75,1.0}
 ]
    \addplot+[thick,solid,
    black!80,
    fill=awesome!60,mark={diamond*},
    boxplot prepared={
        box extend=0.3, 
lower whisker=0.0,
lower quartile=0.0,
median=0.118,
upper quartile=0.221,
upper whisker=1.0
    },
    ] coordinates {};
    \addplot+[thick,solid,
    black!80,
    fill=awesome!60,mark={diamond*},
    boxplot prepared={
        box extend=0.3, 
lower whisker=0.0,
lower quartile=0.076,
median=0.132,
upper quartile=0.204,
upper whisker=1.0
    },
    ] coordinates {};
\nextgroupplot[
    ytick={1,2},
    ylabel=\empty,
    title=\textbf{Affiliate Status},
    xlabel=$\log$ Degree Centrality,  yticklabels={Basic,Affiliate},
    ymin=0.5,
    ymax=2.5,
    xmin=-0.5,
    xmax=8.5,
    xtick={0,2,4,6,8}
 ]

    %DONE
    \addplot+[thick,solid,
    black!80,
    fill=cyan!90,mark={diamond*},
    boxplot prepared={
        box extend=0.3, 
lower whisker=0.0,
lower quartile=2.079,
median=3.091,
upper quartile=4.094,
upper whisker=10.471
    },
    ] coordinates {};
    %DONE
    \addplot+[thick,solid,
    black!80,
    fill=cyan!90,mark={diamond*},
    boxplot prepared={
        box extend=0.3, 
lower whisker=0.0,
lower quartile=2.996,
median=3.761,
upper quartile=4.466,
upper whisker=7.766
    },
    ] coordinates {};

\nextgroupplot[
    ytick={1,2},
    ylabel=\empty,
    title=\textbf{Affiliate Status},
    xlabel=Clustering Coefficient,
    yticklabels={Basic,Affiliate},
    ymin=0.5,
    ymax=2.5,
    xmin=-0.05,
    xmax=1.05,
    xtick={0,0.25,0.5,0.75,1.0},
    xticklabels={0,0.25,0.5,0.75,1.0}
 ]
    \addplot+[thick,solid,
    black!80,
    fill=awesome!60,mark={diamond*},
    boxplot prepared={
        box extend=0.3, 
lower whisker=0.0,
lower quartile=0.067,
median=0.133,
upper quartile=0.214,
upper whisker=1.0
    },
    ] coordinates {};
    \addplot+[thick,solid,
    black!80,
    fill=awesome!60,mark={diamond*},
    boxplot prepared={
        box extend=0.3, 
lower whisker=0.0,
lower quartile=0.082,
median=0.13,
upper quartile=0.198,
upper whisker=1.0
    },
    ] coordinates {};
\nextgroupplot[
    ytick={1,2},
    ylabel=\empty,
    title=\textbf{Explicit Content},
    xlabel=$\log$ Degree Centrality,  yticklabels={Gaming,Mature},
    ymin=0.5,
    ymax=2.5,
    xmin=-0.5,
    xmax=8.5,
    xtick={0,2,4,6,8}
 ]

    %DONE
    \addplot+[thick,solid,
    black!80,
    fill=cyan!90,mark={diamond*},
    boxplot prepared={
        box extend=0.3, 
      median=3.367,
      upper quartile=4.19,
      lower quartile=2.39,
      upper whisker=7.374,
      lower whisker=0.0
    },
    ] coordinates {};
    %DONE
    \addplot+[thick,solid,
    black!80,
    fill=cyan!90,mark={diamond*},
    boxplot prepared={
        box extend=0.3, 
      median=2.996,
      upper quartile=3.67,
      lower quartile=2.30,
      upper whisker=6.24,
      lower whisker=0.00
    },
    ] coordinates {};

\nextgroupplot[
    ytick={1,2},
    ylabel=\empty,
    title=\textbf{Explicit Content},
    xlabel=Clustering Coefficient,
    yticklabels={Gaming, Mature},
    ymin=0.5,
    ymax=2.5,
    xmin=-0.05,
    xmax=1.05,
    xtick={0,0.25,0.5,0.75,1.0},
    xticklabels={0,0.25,0.5,0.75,1.0}
 ]
    \addplot+[thick,solid,
    black!80,
    fill=awesome!60,mark={diamond*},
    boxplot prepared={
        box extend=0.3, 
lower whisker=0.0,
lower quartile=0.124,
median=0.217,
upper quartile=0.319,
upper whisker=1.0
    },
    ] coordinates {};
    \addplot+[thick,solid,
    black!80,
    fill=awesome!60,mark={diamond*},
    boxplot prepared={
        box extend=0.3, 
lower whisker=0.0,
lower quartile=0.148,
median=0.25,
upper quartile=0.379,
upper whisker=1.0
    },
    ] coordinates {};
\end{groupplot}
\end{tikzpicture}
\caption{The box plots of degree centrality and clustering coefficient conditional on the \textit{Dead Account, Affiliate Status}, and \textit{Explicit Content} attributes.}\label{fig:twitch_distribution_others}
\end{figure}
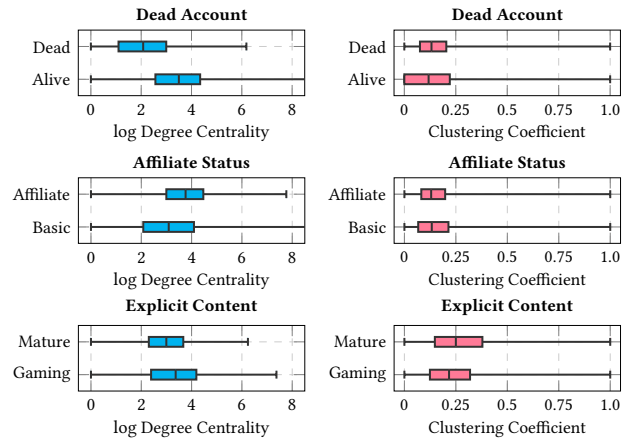

We used boxplots to visualize the distribution of the log transformed degree and clustering coefficient conditional on the categorical vertex attributes. We plotted these boxplots of the structural features on Figures \ref{fig:twitch_distribution_language} and \ref{fig:twitch_distribution_others}. Based on these plots we can deduce that users who broadcast in more commonly spoken languages are well connected. At the same time their friends are less likely to be connected -- this potentially hints at their hub-like role. The results obtained for the other attributes are also intuitive: (i) users who churned from the platform are less central in the social network; (ii) broadcasters who use explicit language are less popular; (iii) those who obtain affiliate status are generally well connected in the social network. These findings hint that all of the categorical features can be embedded with the use of structural role-based node embedding techniques.
\input{./figures/country_box.tex}
\section{Experimental evaluation}\label{sec:twitch_experiments}

We use \textit{Twitch Gamers} to evaluate the predictive value of features extracted with popular node embedding algorithms. The target attributes of node classification were the \textit{Explicit Content}, \textit{Broadcaster Language}, \textit{Dead Account} and \textit{Affiliate Status} variables. In our experiments we used the open-source \textit{Karate Club} \cite{karateclub} library with the default hyperparameter settings of the node embedding procedures. Specifically, we tested the performance of the following proximity preserving node embeddings:

\begin{enumerate}
\item \textbf{Diff2Vec} \cite{rozemberczki2018fast,littleballoffur} factorizes a pointwise mutual information (henceforth PMI) matrix derived from a diffusion process.
\item \textbf{DeepWalk} \cite{deepwalk} decomposes the PMI matrix of summed normalized adjacency matrix powers with implicit factorization.
\item \textbf{Walklets} \cite{perozzidontwalk} factorizes the PMI matrix of normalized adjacency matrix powers to obtain multi-scale node embeddings. 
\item \textbf{RandNE} \cite{zhang2018billion} smooths an orthogonal node embedding matrix with powers of the adjacency matrix.
\end{enumerate}

\begin{figure}[h!]
\centering
\scalebox{0.9}{
\begin{tikzpicture}
\begin{groupplot}[	grid=major,
	grid style={dashed, gray!40},group style={
                      group name=myplot,
                      group size= 2 by 2, horizontal sep=1.55cm,vertical sep=1.2cm},height=3.8cm,width=4.75cm, title style={at={(0.5,0.9)},anchor=south},every axis x label/.style={at={(axis description cs:0.5,-0.15)},anchor=north},]
\nextgroupplot[
 	legend columns=4,
	legend style={at={(1.20,1.25)},anchor=south},
y label style={at={(0.05,0.5)}},
    legend entries={\textbf{Diff2Vec}, \textbf{DeepWalk}, \textbf{Walklets},\textbf{RandNE}},
	ylabel=AUC score,
	xlabel=Training data ratio \%,
	xtick={0.1,0.3,0.5,0.7,0.9},
	xmin=0,
	title=\textbf{Explicit Content},
	xmax=1.0,
	ymin=0.45,
	ymax=1.05,
	ytick={0.5,0.6,0.7,0.8,0.9,1.0},
]
\addplot [very thick, awesome,mark=*,opacity=0.6]coordinates {
(0.1,0.541)
(0.2,0.562)
(0.3,0.593)
(0.4,0.607)
(0.5,0.604)
(0.6,0.614)
(0.7,0.622)
(0.8,0.622)
(0.9,0.625)
};
\addplot [very thick, blue,mark=square*,opacity=0.6]coordinates {
(0.1,0.557)
(0.2,0.572)
(0.3,0.595)
(0.4,0.607)
(0.5,0.62)
(0.6,0.626)
(0.7,0.631)
(0.8,0.633)
(0.9,0.637)
};
\addplot [very thick, ao(english), ,mark=triangle*,opacity=0.6]coordinates {
(0.1,0.614)
(0.2,0.617)
(0.3,0.634)
(0.4,0.642)
(0.5,0.647)
(0.6,0.656)
(0.7,0.659)
(0.8,0.663)
(0.9,0.667)
};
\addplot [very thick, amber,mark=diamond*,opacity=0.6]coordinates {
(0.1,0.528)
(0.2,0.523)
(0.3,0.524)
(0.4,0.525)
(0.5,0.529)
(0.6,0.535)
(0.7,0.538)
(0.8,0.541)
(0.9,0.543)
};

\nextgroupplot[
	xtick={0.1,0.3,0.5,0.7,0.9},
    y label style={at={(0.05,0.5)}},
	ylabel=AUC score,
	xlabel=Training data ratio \%,
	title=\textbf{Broadcaster Language},
	xmin=0,
	xmax=1.0,
	ymin=0.45,
	ymax=1.05,
	ytick={0.5,0.6,0.7,0.8,0.9,1.0},
	]
\addplot [very thick, awesome,mark=*,opacity=0.6]coordinates {
(0.1,0.701)
(0.2,0.744)
(0.3,0.729)
(0.4,0.728)
(0.5,0.725)
(0.6,0.936)
(0.7,0.938)
(0.8,0.937)
(0.9,0.939)
};
\addplot [very thick, blue,mark=square*,opacity=0.6]coordinates {
(0.1,0.763)
(0.2,0.77)
(0.3,0.772)
(0.4,0.76)
(0.5,0.743)
(0.6,0.945)
(0.7,0.945)
(0.8,0.946)
(0.9,0.949)
};
\addplot [very thick, ao(english), ,mark=triangle*,opacity=0.6]coordinates {
(0.1,0.802)
(0.2,0.744)
(0.3,0.745)
(0.4,0.746)
(0.5,0.743)
(0.6,0.963)
(0.7,0.962)
(0.8,0.963)
(0.9,0.965)
};
\addplot [very thick, amber,mark=diamond*,opacity=0.6]coordinates {
(0.1,0.524)
(0.2,0.528)
(0.3,0.513)
(0.4,0.515)
(0.5,0.515)
(0.6,0.75)
(0.7,0.759)
(0.8,0.762)
(0.9,0.773)
};

\nextgroupplot[
	xtick={0.1,0.3,0.5,0.7,0.9},
    y label style={at={(0.05,0.5)}},
	ylabel=AUC score,
	xlabel=Training data ratio \%,
	title=\textbf{Dead Account},
	xmin=0,
	xmax=1.0,
	ymin=0.45,
	ymax=1.05,
	ytick={0.5,0.6,0.7,0.8,0.9,1.0},
	]
\addplot [very thick, awesome,mark=*,opacity=0.6]coordinates {
(0.1,0.538)
(0.2,0.56)
(0.3,0.55)
(0.4,0.546)
(0.5,0.536)
(0.6,0.549)
(0.7,0.575)
(0.8,0.595)
(0.9,0.597)
};
\addplot [very thick, blue,mark=square*,opacity=0.6]coordinates {
(0.1,0.55)
(0.2,0.578)
(0.3,0.595)
(0.4,0.602)
(0.5,0.59)
(0.6,0.607)
(0.7,0.616)
(0.8,0.616)
(0.9,0.62)
};
\addplot [very thick, ao(english), ,mark=triangle*,opacity=0.6]coordinates {
(0.1,0.607)
(0.2,0.699)
(0.3,0.738)
(0.4,0.727)
(0.5,0.75)
(0.6,0.769)
(0.7,0.779)
(0.8,0.78)
(0.9,0.789)
};
\addplot [very thick, amber,mark=diamond*,opacity=0.6]coordinates {
(0.1,0.52)
(0.2,0.517)
(0.3,0.535)
(0.4,0.531)
(0.5,0.533)
(0.6,0.537)
(0.7,0.537)
(0.8,0.533)
(0.9,0.527)
};

\nextgroupplot[
	xtick={0.1,0.3,0.5,0.7,0.9},
    y label style={at={(0.05,0.5)}},
	ylabel=AUC score,
	xlabel=Training data ratio \%,
	title=\textbf{Affiliate Status},
	xmin=0.0,
	xmax=1.0,
	ymin=0.45,
	ymax=1.05,
	ytick={0.5,0.6,0.7,0.8,0.9,1.0},
	]
\addplot [very thick, awesome,mark=*,opacity=0.6]coordinates {
(0.1,0.561)
(0.2,0.609)
(0.3,0.631)
(0.4,0.652)
(0.5,0.663)
(0.6,0.667)
(0.7,0.675)
(0.8,0.677)
(0.9,0.675)
};
\addplot [very thick, blue,mark=square*,opacity=0.6]coordinates {
(0.1,0.569)
(0.2,0.601)
(0.3,0.617)
(0.4,0.643)
(0.5,0.659)
(0.6,0.677)
(0.7,0.681)
(0.8,0.691)
(0.9,0.693)
};
\addplot [very thick, ao(english), ,mark=triangle*,opacity=0.6]coordinates {
(0.1,0.685)
(0.2,0.698)
(0.3,0.704)
(0.4,0.714)
(0.5,0.73)
(0.6,0.736)
(0.7,0.74)
(0.8,0.743)
(0.9,0.746)
};
\addplot [very thick, amber,mark=diamond*,opacity=0.6]coordinates {
(0.1,0.507)
(0.2,0.504)
(0.3,0.51)
(0.4,0.518)
(0.5,0.519)
(0.6,0.525)
(0.7,0.528)
(0.8,0.531)
(0.9,0.529)
};

\end{groupplot}
\end{tikzpicture}}
\caption{Predictive performance of proximity preserving node embedding techniques on classification tasks measured by area under the curve scores on the test set as a function of training set ratio.}\label{fig:twitch_predictive_proximity}
\end{figure}
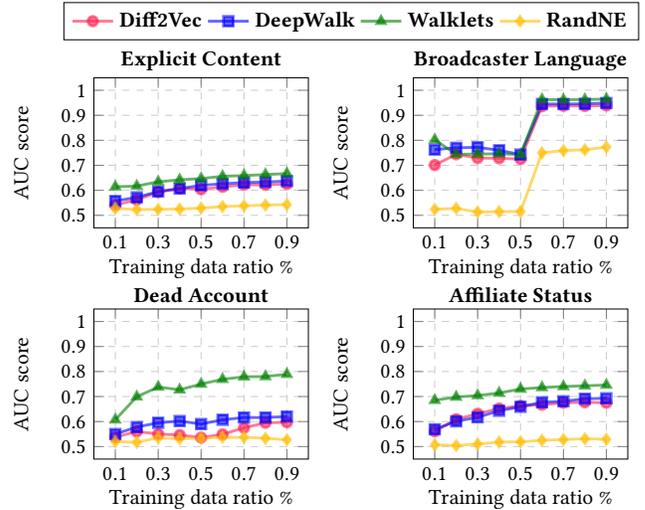

\begin{figure}[h!]
\centering
\scalebox{0.9}{
\begin{tikzpicture}
\begin{groupplot}[	grid=major,
	grid style={dashed, gray!40},group style={
                      group name=myplot,
                      group size= 2 by 2, horizontal sep=1.55cm,vertical sep=1.2cm},height=3.8cm,width=4.75cm, title style={at={(0.5,0.9)},anchor=south},every axis x label/.style={at={(axis description cs:0.5,-0.15)},anchor=north},]
\nextgroupplot[
 	legend columns=4,
	legend style={at={(1.20,1.25)},anchor=south},
y label style={at={(0.05,0.5)}},
    legend entries={\textbf{Role2Vec}, \textbf{ASNE}, \textbf{MUSAE},\textbf{FEATHER}},
	ylabel=AUC score,
	xlabel=Training data ratio \%,
	xtick={0.1,0.3,0.5,0.7,0.9},
	xmin=0,
	title=\textbf{Explicit Content},
	xmax=1.0,
	ymin=0.45,
	ymax=1.05,
	ytick={0.5,0.6,0.7,0.8,0.9,1.0},
]
\addplot [very thick, awesome,mark=*,opacity=0.6]coordinates {

(0.1,0.568)
(0.2,0.567)
(0.3,0.591)
(0.4,0.595)
(0.5,0.599)
(0.6,0.621)
(0.7,0.63)
(0.8,0.636)
(0.9,0.637)
};
\addplot [very thick, blue,mark=square*,opacity=0.6]coordinates {
(0.1,0.564)
(0.2,0.551)
(0.3,0.596)
(0.4,0.599)
(0.5,0.607)
(0.6,0.629)
(0.7,0.64)
(0.8,0.642)
(0.9,0.651)
};
\addplot [very thick, ao(english), ,mark=triangle*,opacity=0.6]coordinates {
(0.1,0.551)
(0.2,0.574)
(0.3,0.591)
(0.4,0.602)
(0.5,0.603)
(0.6,0.61)
(0.7,0.616)
(0.8,0.618)
(0.9,0.624)
};
\addplot [very thick, amber,mark=diamond*,opacity=0.6]coordinates {
(0.1,0.563)
(0.2,0.595)
(0.3,0.617)
(0.4,0.629)
(0.5,0.626)
(0.6,0.634)
(0.7,0.633)
(0.8,0.637)
(0.9,0.64)
};

\nextgroupplot[
	xtick={0.1,0.3,0.5,0.7,0.9},
    y label style={at={(0.05,0.5)}},
	ylabel=AUC score,
	xlabel=Training data ratio \%,
	title=\textbf{Broadcaster Language},
	xmin=0,
	xmax=1.0,
	ymin=0.45,
	ymax=1.05,
	ytick={0.5,0.6,0.7,0.8,0.9,1.0},
	]
\addplot [very thick, awesome,mark=*,opacity=0.6]coordinates {
(0.1,0.778)
(0.2,0.766)
(0.3,0.781)
(0.4,0.772)
(0.5,0.764)
(0.6,0.954)
(0.7,0.954)
(0.8,0.957)
};
\addplot [very thick, blue,mark=square*,opacity=0.6]coordinates {
(0.1,0.81)
(0.2,0.811)
(0.3,0.821)
(0.4,0.815)
(0.5,0.81)
(0.6,0.958)
(0.7,0.958)
(0.8,0.959)
(0.9,0.962)
};
\addplot [very thick, ao(english), ,mark=triangle*,opacity=0.6]coordinates {
(0.1,0.675)
(0.2,0.686)
(0.3,0.701)
(0.4,0.711)
(0.5,0.721)
(0.6,0.769)
(0.7,0.778)
(0.8,0.78)
(0.9,0.784)
};
\addplot [very thick, amber,mark=diamond*,opacity=0.6]coordinates {
(0.1,0.705)
(0.2,0.695)
(0.3,0.7)
(0.4,0.691)
(0.5,0.694)
(0.6,0.864)
(0.7,0.869)
(0.8,0.871)
(0.9,0.869)
};

\nextgroupplot[
	xtick={0.1,0.3,0.5,0.7,0.9},
    y label style={at={(0.05,0.5)}},
	ylabel=AUC score,
	xlabel=Training data ratio \%,
	title=\textbf{Dead Account},
	xmin=0,
	xmax=1.0,
	ymin=0.45,
	ymax=1.05,
	ytick={0.5,0.6,0.7,0.8,0.9,1.0},
	]
\addplot [very thick, awesome,mark=*,opacity=0.6]coordinates {
(0.1,0.59)
(0.2,0.698)
(0.3,0.722)
(0.4,0.712)
(0.5,0.721)
(0.6,0.727)
(0.7,0.718)
(0.8,0.7)
(0.9,0.712)
};
\addplot [very thick, blue,mark=square*,opacity=0.6]coordinates {
(0.1,0.674)
(0.2,0.696)
(0.3,0.722)
(0.4,0.702)
(0.5,0.729)
(0.6,0.745)
(0.7,0.757)
(0.8,0.761)
(0.9,0.771)
};
\addplot [very thick, ao(english), ,mark=triangle*,opacity=0.6]coordinates {
(0.1,0.547)
(0.2,0.586)
(0.3,0.6)
(0.4,0.591)
(0.5,0.661)
(0.6,0.688)
(0.7,0.707)
(0.8,0.731)
(0.9,0.733)
};
\addplot [very thick, amber,mark=diamond*,opacity=0.6]coordinates {
(0.1,0.625)
(0.2,0.586)
(0.3,0.565)
(0.4,0.588)
(0.5,0.598)
(0.6,0.597)
(0.7,0.639)
(0.8,0.671)
(0.9,0.661)};

\nextgroupplot[
	xtick={0.1,0.3,0.5,0.7,0.9},
    y label style={at={(0.05,0.5)}},
	ylabel=AUC score,
	xlabel=Training data ratio \%,
	title=\textbf{Affiliate Status},
	xmin=0,
	xmax=1.0,
	ymin=0.45,
	ymax=1.05,
	ytick={0.5,0.6,0.7,0.8,0.9,1.0},
	]
\addplot [very thick, awesome,mark=*,opacity=0.6]coordinates {
(0.1,0.654)
(0.2,0.659)
(0.3,0.673)
(0.4,0.68)
(0.5,0.691)
(0.6,0.698)
(0.7,0.697)
(0.8,0.699)
(0.9,0.697)
};
\addplot [very thick, blue,mark=square*,opacity=0.6]coordinates {
(0.1,0.655)
(0.2,0.649)
(0.3,0.668)
(0.4,0.689)
(0.5,0.698)
(0.6,0.705)
(0.7,0.716)
(0.8,0.724)
(0.9,0.725)
};
\addplot [very thick, ao(english), ,mark=triangle*,opacity=0.6]coordinates {
(0.1,0.658)
(0.2,0.686)
(0.3,0.688)
(0.4,0.695)
(0.5,0.698)
(0.6,0.703)
(0.7,0.71)
(0.8,0.712)
(0.9,0.713)
};
\addplot [very thick, amber,mark=diamond*,opacity=0.6]coordinates {
(0.1,0.654)
(0.2,0.644)
(0.3,0.664)
(0.4,0.676)
(0.5,0.683)
(0.6,0.684)
(0.7,0.688)
(0.8,0.69)
(0.9,0.693)
};

\end{groupplot}
\end{tikzpicture}}
\caption{Predictive performance of structural role preserving node embedding techniques on classification tasks measured by area under the curve scores on the test set as a function of training set ratio.}\label{fig:twitch_predictive_structural}
\end{figure}
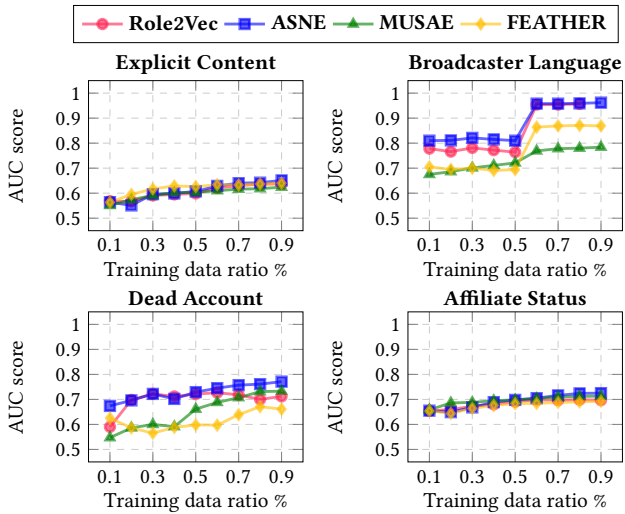

We evaluated the value of features extracted with these structural role-based node embedding algorithms:
\begin{enumerate}
    \item \textbf{Role2Vec} \cite{ahmed2018learning} decomposes the PMI matrix node -- tree feature co-occurrences with an implicit factorization technique.
    \item \textbf{ASNE} \cite{asne} factorizes a target matrix obtained by concatenating the adjacency matrix and a structural feature matrix which includes one-hot encodings of the log degree and clustering coefficient.
    \item \textbf{MUSAE} \cite{musae} learns multi-scale structural role-based node embeddings from matrices obtained by multiplying the structural feature matrix with adjacency matrix powers. 
    \item \textbf{FEATHER} \cite{feather} distills node embeddings from graph characteristic functions of the log transformed degree and clustering coefficient. 
\end{enumerate}
We used the \textit{scikit-learn} \cite{scikit,scikitapi} implementation of logistic regression with the default hyperparameter settings to predict the node labels using the node embeddings as input features. It has to be noted that these default settings involve the use of weight regularization, because of this each node embedding dimension was normalized. The classifiers were trained with various highly skewed train/test data split ratios by  utilizing less than 1\% of training data. We plotted mean macro-averaged AUC scores on the test set calculated from 10 random seed train/test splits on Figures \ref{fig:twitch_predictive_proximity} and \ref{fig:twitch_predictive_structural}. 

The most important finding based on our results is that the target attributes in \textit{Twitch Gamers} are suitable for testing the predictive power of features extracted with both proximity preserving and structural role-based node embeddings. These results showcase that certain node embedding techniques have a considerable advantage on the downstream tasks. We also see evidence that proximity preserving algorithms extract features which are more useful for predicting \textit{Broadcaster Language}. This was expected based on our empirical analysis as it is an attribute which most probably strongly influences linking behaviour. Another similar intuitive finding is that structural role-based embedding techniques\cite{ahmed2018learning} create more expressive features for predicting the \textit{Dead Account} target variable. This is not surprising, our descriptive analysis had shown that users who churned from the platform have idiosyncratic structural attributes. Our results also verify the known fact that multi-scale proximity preserving node embeddings, such as \textit{Walklets} \cite{perozzidontwalk} and \textit{GraRep} \cite{cao2015grarep}, outperform techniques like \textit{DeepWalk} \cite{deepwalk} that pool information form low and higher order proximities. 
\section{Conclusions}\label{sec:twitch_conclusions}
We introduced \textit{Twitch Gamers} a medium sized social network dataset with a rich set of potential target attributes. Our descriptive analysis of the dataset had demonstrated that both proximity preserving and structural role-based node embeddings can potentially distill high quality features for node classification. We verified this precognition by a series of experiments. Our findings show that \textit{Twitch Gamers} can serve as an important benchmark to assess novel node embedding techniques. We are particularly excited that the prediction of certain vertex attributes turned out to be challenging machine learning task.

\section*{Acknowledgements}
Benedek Rozemberczki was supported by the Centre for Doctoral Training in Data Science, funded by EPSRC (grant EP/L016427/1).

\bibliographystyle{ACM-Reference-Format}

\bibliography{main}
\typeout{get arXiv to do 4 passes: Label(s) may have changed. Rerun}
\end{document}